\documentclass[twocolumn,aps,amsmath,english,superscriptaddress,longbibliography,notitlepage]{revtex4-1}
\usepackage{graphicx,natbib,subfigure,tabularx,epsfig,longtable,amsfonts,rotating,mathrsfs}
\usepackage{color,babel,currfile,subfigure,hyperref}
\usepackage[normalem]{ulem}
\usepackage{float}
\usepackage{array}
\begin{document}
\title{Dynamics Reflect Gapless Edge Modes for Topological Superconductor}
\date{\today}

\author{X.L. Zhao}
\affiliation{Graduate School of China Academy of Engineering Physics, China\\}
\affiliation{Quantum Physics Laboratory,  School of Science, Qingdao Technological University, Qingdao,
266033, China\\}

\author{M. Li}
\affiliation{Qingdao University of Technology, 0532, Qingdao, Shandong, China}

\author{T.H. Qiu}
\affiliation{Qingdao University of Technology, 0532, Qingdao, Shandong, China}

\author{L.B. Chen}
\affiliation{Quantum Physics Laboratory,  School of Science, Qingdao Technological University, Qingdao,
266033, China\\}

\author{Y.L. Ma}
\affiliation{Qingdao University of Technology, 0532, Qingdao, Shandong, China}

\author{H.Y. Ma$^*$}
\affiliation{Qingdao University of Technology, 0532, Qingdao, Shandong, China}

\author{X.X. Yi$^*$}
\affiliation{Center for Quantum Sciences and School of Physics, Northeast Normal
University, Changchun 130024, China}

\begin{abstract}
We focus on the dynamical feature for a $p$-wave superconductor model in different
parameter regions in terms of the appearance of gapless edge modes in reel
geometry. Firstly, we show the parameter region with gapless edge modes versus
a parameter and quasi-momentum. The parameter diagram can be reflected by the
expectations of Pauli matrices in global manners. In another view, the dynamical
feature of the excitation behave differently in the parameter regions with
topological gapless edge modes and not. And the cusps of dynamical return rate
vanish as the parameter pass the boundary in the parameter region slowly enough.
It is found that the dynamics in the parameter region with gapless edge modes
behaves differently to that without edge modes and related mostly to the
eigenenergy gap between the pre-and post-quench eigenstates. The cusps of the
dynamical return rate behave robustly against the noise in the lattice until
localization behavior dominates. This work benefits detecting topological edge
modes by dynamical manners.
\end{abstract}
\maketitle
\section{Introduction}
Different matter in one phase belong to the same equivalence class according
to certain criteria. Distinct from Landau's theory about describing phase of
matter based on order parameter, topological invariants act as new criteria
in distinguishing phases of matter, like insulators, semimetals, and
superconductors by a novel mechanism \cite{JPC61181,PRL49405,PRL501153}.
The topological $p$-wave superconductors which support gapless edge modes,
are crucial for their implementation as ingredients for quantum-computing
devices~\cite{PRB6110267,PU44131,PRL86268,RPP7507501,TITS}, and have
attracted wide attention~\cite{PRL103020401,PRL104040502,PRB81125318}. Such
superconductors can be realized by depositing magnetic atoms on a conventional
$s$-wave superconductor substrate~\cite{Science346602,NQI216035}. The topological
phase can be characterized by topological invariants~\cite{AP313497,PRL95146802}.
We would focus on a topological superconductor model in this work.

Nonequilibrium phase transition is another novel topic attracting intensive
attention which is also not based on the mechanism of Landau's order parameter
theory about phase of matter. Such a phase transition occurs when quenching
across the critical time for a transverse field Ising model~\cite{PRL110135704}
and quantum many-body spin system~\cite{PRL119080501,PRL124043001}. These works
inspire us to check the dynamical feature versus the appearance of topological
gapless edge modes.

The homotopy invariant of loop constructed from time-evolution operator can
represent the dynamical topology and opens up a novel avenue to classify quench
dynamics~\cite{PRL124160402}. Dynamically topological order is found closely
related to the singularity of the Bogoliubov angle, but further investigation
is needed for the quenches in the view of topology~\cite{PRB102060409}.
Topology-changing quenches have been investigated followed by dynamical phase
transitions for various topological models~\cite{PRB91155127}. The dynamical
feature of localization-delocalization transition for a one-dimensional
incommensurate lattice described by the Aubry-Andr$\acute{e}$ model has been
investigated by quenching~\cite{PRB95184201}. This inspires us checking the
localization behavior during dynamics in this work.

In this work, we consider a topological superconductor model of reel geometry
originated from Kitaev's model~\cite{PU44131}. Firstly, the parameter region
for the appearance of gapless edge modes is shown. The appearance of edge modes
is also reflected by the expectations of Pauli matrices in global manners.
Secondly, the dynamical properties in terms of gapless edge modes are checked
including dynamical return rate, dynamics of excitations, and localization during
evolution.

\begin{figure*}[tbp]
\includegraphics[width=18cm]{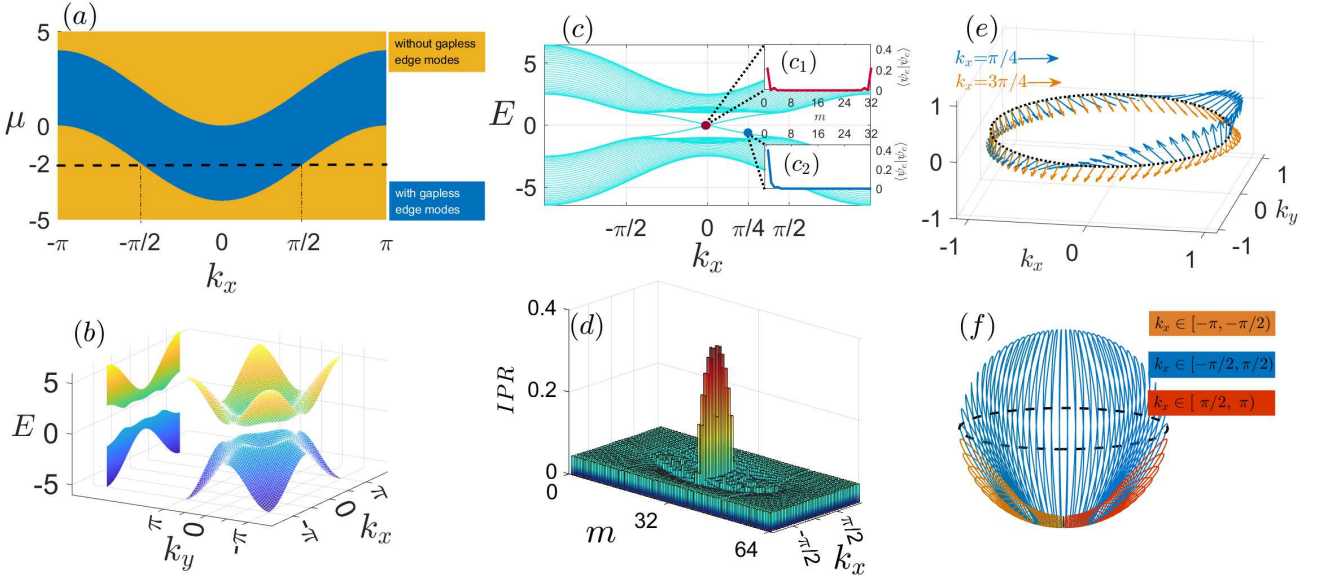}
\caption{($a$) The parameter diagram drew according to Eq.(\ref{phasediagram}),
indicates the regions for the appearance of gapless edge modes for the reel-shape
topological superconductor with periodic and open boundary conditions in $x$ and
$y$ directions, respectively. The blue region denotes that with gapless edge
modes and the other regions denote those without gapless edge modes. ($b$)
Dispersion versus $k_x$ and $k_y$ with periodic boundary conditions in both
$x$ and $y$ directions. ($c$) Branches of the dispersion relation versus $k_x$
when $\mu=-2$, which is the main research situation in this work. The insets
($c_1$) and ($c_2$) show the distributions $\langle\psi_e |\psi_e\rangle$ for
the edge modes $|\psi_e\rangle$ (half of the eigenstate is shown due to redundancy
of the Hilbert space). ($d$) $IPR$ for the eigenstates (numbered by $m$) versus
$k_x$ when $\mu=-2$. ($e$) The vector of
$(\langle\sigma_x\rangle,\langle\sigma_y\rangle,\langle\sigma_z\rangle)$
for two examples in different parameter regions ($k_x=\pi/4$ and $k_x=3\pi/4$)
when $\mu=-2$ for a full period of $k_y$. ($f$) A range of trajectories for
$(\langle\sigma_x\rangle,\langle\sigma_y\rangle,\langle\sigma_z\rangle)$
when the trajectories are drew of $k_y\in[-\pi,\pi)$ for three regions of
$k_x\in[-\pi,-\pi/2)$, $[-\pi/2,\pi/2)$, $[\pi/2,\pi)$ with different dominant
colors (slightly gradient color for different $k_x$).}
\label{EnergySpKeps}
\end{figure*}

Following in Sec.~\ref{Model}, the $p$-wave superconductor model which we focus on
is introduced. In Sec.~\ref{Phasediagram}, the parameter diagram in terms of the
appearance of gapless edge modes and the relation with expectation values of Pauli
matrices are studied. In Sec.~\ref{dynamicalaspect}, the dynamical feature for this
model is checked. Finally, we conclude in Sec. \ref{CoC}.
\section{The Topological Superconductor Model}
\label{Model}
We consider a two-dimensional topological $p$-wave superconductor model at the
mean-field level described by the lattice Hamiltonian
\begin{eqnarray}
\begin{split}
H=&\sum\limits_{n_x,n_y} -\mu a_{n_x,n_y}^{\dagger}a_{n_x,n_y}\\
&- \kappa(a_{n_x,n_y}^{\dagger}a_{n_x+1,n_y}+a_{n_x,n_y}^{\dagger}a_{n_x,n_y+1}+h.c.)\\
&+(\lambda_xa^{\dagger}_{n_x,n_y}a^{\dagger}_{n_x+1,n_y}+\lambda_ya^{\dagger}_{n_x,n_y}a^{\dagger}_{n_x,n_y+1}+h.c.) ,\\
\label{HamiltonianR}
\end{split}
\end{eqnarray}
where $n_x$ and $n_y$ denote the positions for the sites in $x$ and $y$ directions
on the lattice, respectively. The operator $a_{n_x,n_y}$ $(a_{n_x,n_y}^{\dag})$
annihilates (creates) a fermion on the lattice site with the coordinate $(n_x,n_y)$.
$\mu$ is the chemical potential. Such a quantity in one dimensional topological
superconductor corresponds to external magnetic field when it is
transformed to the transverse-field $XY$ model by Jordan-Wigner transformation
~\cite{PRA21075,PRL107010403}. $\kappa$ is the hopping matrix element
used as the unit in this work ($\kappa$=1) and $\lambda_{x(y)}$=$\lambda_0e^{i\theta_{x(y)}}$.
We focus on the case $\lambda_0$=$1/2$, $\theta_{x(y)}$=$0(\pi/2)$ in this work. Namely,
the pairing amplitude is anisotropic with an additional phase of $\pi/2$ in $y$-direction
compared to $x$-direction. It is a spin-polarized fermion model with the spin degree
of freedom frozen by magnetic field.

After the Fourier transformation in $x$ and $y$ directions, we obtain the Hamiltonian
in the quasi-momentum space
\begin{equation}
\begin{aligned}
\mathcal{H}&=\sum_{\alpha=x,y,z}h_{\alpha}\sigma_{\alpha}\\
&=\sum_{k_x,k_y}(\sin k_y- i \sin k_x)\sigma_++h.c.+\xi_k\sigma_z,
\end{aligned} \label{Hkk}
\end{equation}
where $\sigma_{\pm}$=$(\sigma_x\pm i\sigma_y)/2$. $\sigma_{x,y,z}$ are Pauli matrices
and $\xi_k$=$-2(\cos k_x+\cos k_y)-\mu$ with $4\kappa$ shifted without loss of generality.
Note that $\mathcal{H}$ has particle-hole symmetry but no time-reversal symmetry.
It is a two-dimensional representation of symmetry class-$D$ describing $Z$ topological
superconductor according to the 10-fold classification~\cite{PRB551142,PRB78195125}.
Due to geometrical symmetry, we consider the periodic and open boundary conditions in
$x$ and $y$ directions, respectively. This is a reel model closing a strip lattice to
a cylinder surface with only Fourier transformation taken in $x$ direction. Then the
Hilbert space is redundantly expanded by the particle-hole basis $\phi^{\dag}_{k_x}$=
$(a^{\dag}_{k_x,1},a_{-k_x,1},a^{\dag}_{k_x,2},a_{-k_x,2},...,a^{\dag}_{k_x, N_y},a_{-k_x, N_y})$,
where $a_{k_x,n_y}=\frac{1}{\sqrt{2N_x}}\sum_{n_x=1}^{N_x}a_{n_x,n_y}e^{ik_xn_x}$.
We would focus on the parameter diagram about the appearance of gapless edge modes,
and the corresponding dynamical feature for this model.

\section{The Parameter Diagram}
\label{Phasediagram}
In this work, we take periodic boundary condition along $x$ direction and open boundary
condition along $y$ direction. Then $k_x$ acts as a good quantum number. The topological
phase can be reflected by integrals~\cite{PRL512250,PRB74085308,PRL101186805}. For example,
the integral in the Brillouin zone (FBZ):
$\sigma_{xy}=\frac{1}{4\pi}\iint_{FBZ}\hat{h}\cdot(\partial_{k_x}\hat{h}\times\partial_{k_y}\hat{h})dk_xdk_y$,
where $\hat{h}$ is the normalized vector of $(h_x,h_y,h_z)$ in the Eq.(\ref{Hkk}) \cite{PRB74085308}.
After some algebra, one can find $\sigma_{xy}=\frac{1}{4\pi}\iint_{FBZ}\hat{n}\cdot(\partial_{k_x}\hat{n}\times\partial_{k_y}\hat{n})dk_xdk_y$,
where $\hat{n}$ is the normalization for $(\langle\sigma_x\rangle,\langle\sigma_y\rangle,\langle\sigma_z\rangle)$
under the ground state~\cite{PRL512250,PRL101186805}. These calculations can be examnied
by using the ground state in the eigen-equations for the Hamiltonian (\ref{Hkk}) as below
\begin{equation}
\mathcal{H}|\pm\rangle=\pm\Omega|\pm\rangle,
\label{EQ}
\end{equation}
where $\Omega=\sqrt{h_{x}^2+h_{y}^2+h_{z}^2}$, and
\begin{eqnarray}
|\pm\rangle=\frac{\pm1}{\sqrt{2\Omega(\Omega\pm h_z)}}\left [
\begin{array}{ll}
h_x-ih_y \\
-h_z\mp\Omega, \\
\end{array}
\right ].
 \label{psi}
\end{eqnarray}
After calculating $\sigma_{xy}$, it can be verified that the topological
nontrivial parameter interval is $\mu\in(-4,4)$.

In topological classification of matter, bulk-boundary correspondence is
a remarkable characteristic. The appearance of topological gapless edge modes
is a signature for topological phase. As we focus on reel-shape lattice, we
pay special attention on the parameter region with the appearance of the
gapless edge modes versus $\mu$ and $k_x$ as shown in FIG.~\ref{EnergySpKeps}(a).
We found that the gapless edge modes locate between the parameter boundaries
described by the relations:
\begin{equation}
\mu=-2\cos k_x \pm2.
\label{phasediagram}
\end{equation}
The region for the appearance of the topological gapless edge modes is coincident
with FIG.~\ref{EnergySpKeps}(a) shown by a cartoon in the supplementary material~\cite{movEq3}.

To check the details of the topological edge modes, the eigenenergy spectra in
quasi-momentum space (with Fourier transformation taken only in $x$ and $y$ directions)
and a reel-shape geometry (with Fourier transformation taken only in $x$ direction)
are shown in FIG.~\ref{EnergySpKeps} (b) and (c), respectively. The subfigures
FIG.~\ref{EnergySpKeps} ($c_1$) and FIG.~\ref{EnergySpKeps}($c_2$) show the halves
for the two distributions of the edge modes.

The topological edge modes have remarkable localization characteristic compared to
ordinary modes. The localization for an state can be revealed by the quantity
$IPR$=$\sum_j^{N_y}|\psi_j^{m,k_x}|^4$, here $\psi_j^{m,k_x}$ denotes the projection
of the $m$th eigenstate on the $j$th site for certain $k_x$ \cite{PR1393}. The
localization for the eigenstates identified by $m$ versus $k_x$ are shown in
FIG.~\ref{EnergySpKeps}(d). It can be seen that the region for the appearance of
the gapless edge modes is consistent with the parameter diagram in
FIG.~(\ref{EnergySpKeps})(a) and (c).

Topological phase of matter is usually indicated by the invariants like Zak
phase, Chern number and so on. In this work, the expectations of Pauli matrices
under ground state, namely, $\langle\sigma_x\rangle$, $\langle\sigma_y\rangle$,
and $\langle\sigma_z\rangle$ can also be applied to distinguish different parameter
regions in terms of gapless edge modes. For example, the vector of
$(\langle\sigma_x\rangle,\langle\sigma_y\rangle,\langle\sigma_z\rangle)$ plays
such a role for this model shown in FIG.~\ref{EnergySpKeps}(e). A reference loop
is parameterized as $(\cos k_x,\sin k_y)$. The vectors wind this loop a full
round when there is topological gapless edge modes ($k_x=\pi/4$ as an example)
whereas not when there is no gapless edge modes ($k_x=3\pi/4$ as an example),
obtained for $k_y$ being ergodic $(-\pi,\pi]$. This agrees with
FIG.~\ref{EnergySpKeps}(a) and (c). Besides, the trajectories of
$\langle\sigma_x\rangle-\langle\sigma_y\rangle-\langle\sigma_z\rangle$ form
different loops in terms of the appearance of gapless edge modes on the Bloch
sphere as shown in FIG.~\ref{EnergySpKeps}(f). These results suggest that the
appearance of topological gapless edge modes can be revealed by expectation
values of operators in global manners. This benefits checking the topological
phase of matter in experiments.
\section{Dynamical Feature}
\label{dynamicalaspect}
Since the appearance of gapless edge modes is a remarkable signature for
topological phase of matter and the structure of energy band is related to
the dynamics of the excitations in lattices. One may conjecture that the
presence of gapless edge modes may be reflected by the dynamics of
excitations. Thus, we check the dynamical features versus the appearance of
gapless edge modes, including quenching by turning the parameter $\mu$
suddenly and the evolution of the excitation in different parameter regions
in the following.
\subsection{Quenching}
\label{DPT}
In conventional thermodynamics and statistical physics, non-analyticities of free
energy density at critical temperatures indicate phase transition in thermodynamic
limit. Similarly, dynamical phase transition describes the analog behavior occurs
in quantum systems when time evolution acts as temperature variation with
non-analytic behavior during evolution. Quenching is a candidate to reflect the
dynamical phase transition by instantaneously changing the parameter(s) of a
Hamiltonian starting from a ground state~
\cite{PRL110135704,PRB102060409,PRB91155127}. The Loschmidt overlap
defined as
\begin{equation}
G(t)=|\langle \psi_{\mu;0} | e^{-i H^{\prime} t} |\psi_{\mu;0} \rangle|,
\label{GHT}
\end{equation}
is a quantity applied in quench. $|\psi_{\mu;0}\rangle$ denotes the initial state
with parameter $\mu$ in the Hamiltonian. This quantity measures the overlap between
the time evolved state $e^{-i H^{\prime} t} |\psi_{\mu;0} \rangle$ and the initial
state $|\psi_{\mu;0}\rangle$ following a sudden change of the parameter $\mu$ in
the post-quench Hamiltonian, namely, $H(\mu)\underrightarrow{~quench~}H(\mu^{\prime})$
in this work.

\begin{figure}[tbp]
\includegraphics[width=8.5cm]{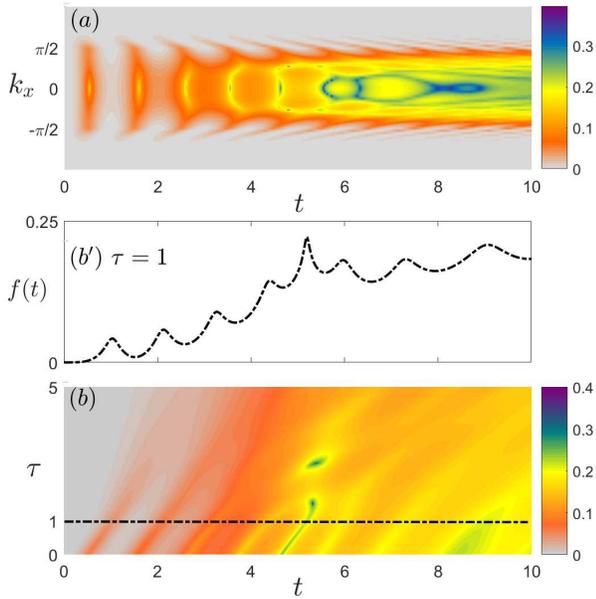}
\caption{($a$) $f(t)$ as a function of time $t$ and $k_x$ when $\mu=-2$. ($b$) $f(t)$
versus time $t$ and $\tau$ introduced in Eq.(\ref{Tau}) when $k_x$=$\pi/4$. ($b^{\prime}$)
shows $f(t)$ versus time when $\tau=1$ corresponding to the dash-dotted line in ($b$).}
\label{Quencheps}
\end{figure}

In statistical physics, the zeros of a partition function correspond to the cusps
of free energy density in thermodynamic limit as a function of temperature. Similarly,
the Loschmidt overlap in Eq.(\ref{GHT}) plays the role as the partition function~
\cite{PRL110135704}. Since $G(t)$ scales with the size of the system $N$,
another quantity, return rate is usually employed in quenching
\begin{equation}
f(t)=-\frac{1}{N} \ln G(t) \,.
\label{eqdyn}
\end{equation}
\cite{PRL110135704,PozsgayLecho}. This quantity behaves non-analytically versus
time when the Loschmidt overlap $G(t)$ vanishes. We employ this quantity to check
the dynamics in terms of the appearance of gapless edge modes. $f(t)$ is shown
versus time $t$ and $k_x$ in FIG.~\ref{Quencheps}(a) when $\mu$ changes suddenly
from $-2$ to $-5$. While $\mu=-2$ and $k_x\in(-\pi/2,\pi/2)$ in the Brillouin zone,
the gapless edge modes appear, and when $k_x$ belongs to the complementary set,
the gapless edge modes disappear, as shown in FIG.~\ref{EnergySpKeps} (c). While
$\mu=-5$, this model is in the topologically trivial phase in the whole Brillouin
zone without gapless edge modes. As can be seen in FIG.~\ref{Quencheps}(a), the
behavior of $f(t)$ is different obviously whether or not $\mu$ quenching across
the parameter boundary of the regions with and without gapless edge modes.

Besides the sudden change of parameters in quench, it is natural to consider
what happens if the parameter $\mu$ continuously crossing such parameter boundaries
within finite time. Among various manners of the parameter crossing the boundaries,
we consider the case
\begin{eqnarray}
\mu(t)=\left \{
\begin{array}{ll}
 \mu_0+\frac{(\mu_f-\mu_0)t}{\tau}~~~&t\leq\tau, \\
\mu_f~~~&t>\tau, \\
\end{array}
\right.
 \label{Tau}
\end{eqnarray}
$\mu_f=-5$ and $\mu_0=-2$ in this work. The larger $\tau$ is, the further it from
quenching and the closer to adiabatic process. The results are shown in
FIG. \ref{Quencheps} (b). It can be seen that the cusps appear and delay until
$\tau$ becomes too large. The slower $\mu(t)$ crossing the parameter boundaries,
the latter the cusps appear since it takes time for reaching the boundary.
Different response for the quench across the parameter boundary may provide the
reference to design control strategies to finish certain research for this model.
\begin{figure}[tbp]
\includegraphics[width=9cm]{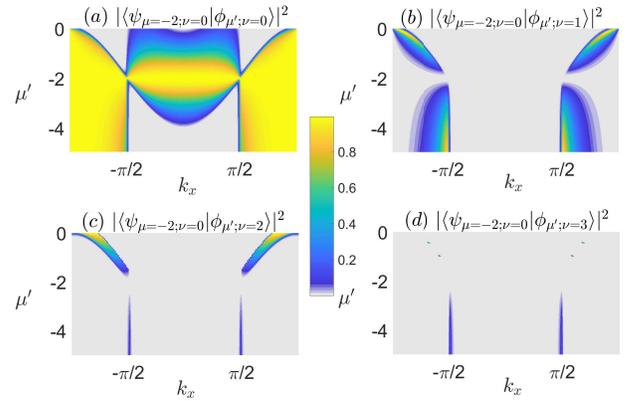}
\caption{($a$)-($d$) $|\langle\psi_{\mu=-2;\nu=0}|\phi_{\mu^{\prime};\nu}\rangle|^{2}$
as a function of $k_x$ and $\mu^{\prime}$ for $\nu$=0,1,2,3, respectively. $|\psi_{\mu=-2;\nu=0}\rangle$
is the ground state of the pre-quench Hamiltonian and $|\phi_{\mu^{\prime};\nu}\rangle$ are the
eigenstates of the post-quench Hamiltonian when the chemical potential is $\mu^{\prime}$. The eigenstates
with larger $\nu$ denote those with eigenvalues further from the one corresponding to $|\psi_{\mu=-2;\nu=0}\rangle$.}
\label{kxkyE}
\end{figure}
\subsection{Overlap between Eigenstates}
\label{OBE}
We conjecture the overlap between the eigenstates of the post-quench Hamiltonian
and the initial state (pre-quench) ($|\psi_{\mu=-2;\nu=0}\rangle$) is a critical
factor influencing the appearance of the cusps during the evolution. Here $\nu$
is the number for different eigenstates. Thus we check the projection
$|\langle\psi_{\mu=-2;\nu=0}|\phi_{\mu^{\prime};\nu}\rangle|^{2}$, where $|\phi_{\mu^{\prime};\nu}\rangle$
are the eigenstates of the post-quench Hamiltonian with the parameter $\mu$ having
changed to $\mu^{\prime}$ abruptly. This projection measures the distance between
$|\psi_{\mu=-2;\nu=0}\rangle$
and $|\phi_{\mu^{\prime};\nu}\rangle$. The larger $\nu$ corresponds to the larger
eigenenergy gap between the corresponding eigenenergies of
$|\psi_{\mu=-2;\nu=0}\rangle$ and $|\phi_{\mu^{\prime};\nu}\rangle$. The results
are shown in FIG.~\ref{kxkyE}. Comparing FIG.~\ref{kxkyE} (a) with FIG.~\ref{EnergySpKeps} (a),
the projection $|\langle\psi_{\mu=-2;\nu=0}|\phi_{\mu^{\prime};\nu=0}\rangle|^{2}$
approaches zero when $\mu$ changes from the region with (without) gapless edge
modes to that without (with) such modes. By comparing FIG.\ref{kxkyE} (a)-(d),
when it quenches to the same parameter region, the distribution of
$|\langle\psi_{\mu=-2;\nu=0}|\phi_{\mu^{\prime};\nu}\rangle|^{2}$ take largest
values when $\nu=0$. Since larger $\nu$ denotes the eigenstate with an eigenvalue
further from the one corresponding to $|\psi_{\mu=-2;\nu=0}\rangle$. Those
distributions of $|\langle\psi_{\mu=-2;\nu=0}|\phi_{\mu^{\prime};\nu}\rangle|^{2}$
with larger $\nu$ show weaker correlation to whether it quenches to different
parameter regions. Thus, it means the eigenstate with the nearest eigenenergy
in the post-quench Hamiltonian is correlated prominently to the cusps in the
dynamics.
\begin{figure}[tbp]
\includegraphics[width=8.5cm]{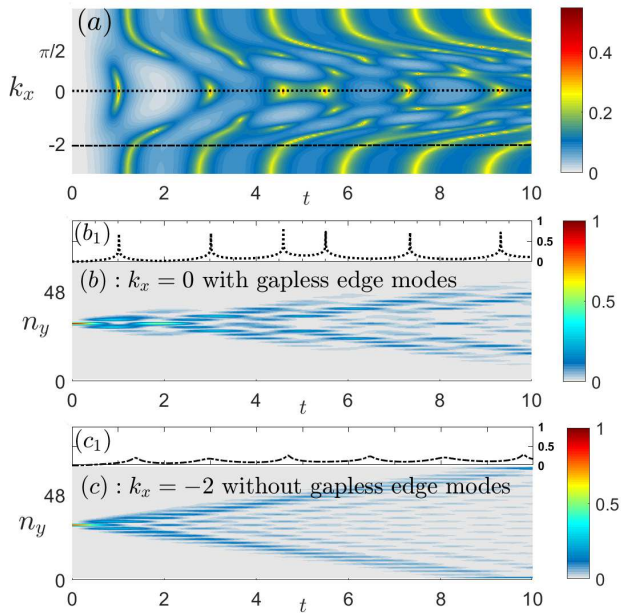}
\caption{($a$) $f^{\prime}(t)$ versus time $t$ and $k_x$ starting from the initial
state with all population on the middle of the lattice. The other parameters in the
Hamiltonian are same to those in FIG.~\ref{Quencheps}. ($b$) and ($c$) show the
projections of the excitation on the lattice during evolution when $k_x$=0 (with
gapless edge modes) and $-2$ (without gapless edge modes), respectively. And a full
examination of $k_x$ in the Brillouin zone is in the supplementary material~\cite{SMV}.
($b_1$) and($c_1$) show the corresponding $f^{\prime}(t)$ for ($b$) and($c$),
respectively.}
\label{ESEvolutioneps}
\end{figure}
\subsection{Dynamical Features in Different Parameter Regions}
\label{Excitation}
The dynamical feature of excitations may behave distinctively in different parameter
regions in terms of the appearance of gapless edge modes. To examine this, we employ
a quantity $f^{\prime}(t)$ with the definition formally like the return rate $f(t)$,
and the projection of the excitation on the lattice site during evolution in
different parameter regions in FIG.~\ref{ESEvolutioneps}. It should be emphasized
that, here we do not use an eigenstate of the pre-quench Hamilonian as the initial
state in $f^{\prime}(t)$, but the excitation with full population on one site as the
initial state. As shown in FIG.~\ref{ESEvolutioneps} ($b$) and ($c$), the free
dynamics of the excitation behaves distinctively in different parameter regions
in terms of edge modes. And a full range of examination is in the supplementary
material~\cite{SMV}. The results hint that the dynamics of the excitation provides
a candidate to denote parameter regions with gapless edge modes or not.
\subsection{Localization Resulting from Noise}
\label{TimeAnderson}
\begin{figure}[tbp]
\includegraphics[width=8.5cm]{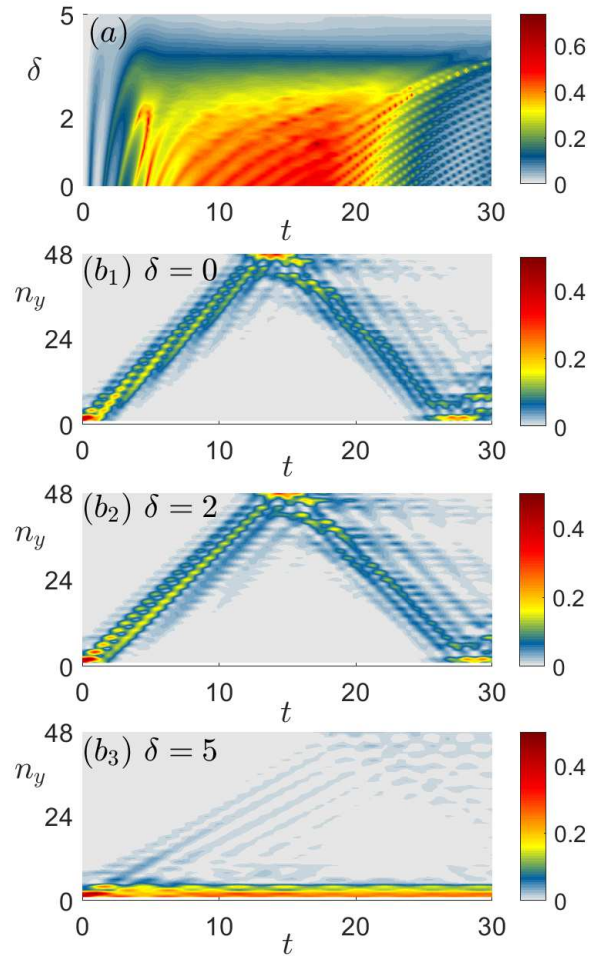}
\caption{($a$) $f(t)$ versus the amplitude $\delta$ of the noise and time $t$
when $N_y$=24 and $k_x$=$\pi/4$. The other parameters are same to those in
FIG.~\ref{Quencheps}. ($b_1$)-($b_3$) show the dynamics of the excitations when
the noise amplitude $\delta$=0, 2 and 5, respectively.}
\label{QuenchNoiseeps}
\end{figure}
Noise and disorders are usually inevitable in reality. The robust property
against disorders of topological edge modes makes them be potential ingredients
for quantum computation. As mentioned previously, the appearance of the gapless
edge modes is related to the appearance of cusps in $f(t)$. We explore the
behavior of $f(t)$ in the presence of the time-dependent noise in $\mu$ after
quenching from the region with gapless edge modes to that without gapless edge
modes. The noise is introduced as $\mu_{\delta}(t)$=$\mu+\delta\mu(t)$, where
$\delta\mu(t)$ denotes the fluctuation with values randomly distributed in
$[0,\delta]$ during evolution. Different from the noise in thermodynamics and
statistical physics which denotes the thermal vibration of the particles and
leads to thermalization, this noise results from the random vibration of the
lattice. In FIG.~\ref{QuenchNoiseeps} (a), we show the dynamics of $f(t)$ versus
$\delta$ and time. It can be seen that cusps occurs until the noise makes
$\mu_{\delta}(t)$ cross the parameter boundary between regions with and without
gapless edge modes as shown in FIG.~\ref{QuenchNoiseeps} $(a)$.

To check the influence of the noise on the dynamics in detail, three examples
of the dynamics quenching from the gapless edge modes are shown in
FIG.~\ref{QuenchNoiseeps} $(b_1)-(b_3)$. The dynamics is depressed, and
localization occurs as the excitation is mainly bounded near the edge with
$\delta$ increasing. Meanwhile, the cusps in $f(t)$ vanish meanwhile the
dynamical pattern changes to those in the region without gapless edge modes,
like that in FIG.~\ref{ESEvolutioneps} $(c)$.

Disorders in real space lead to localization of quantum states due to Anderson
localization mechanism~\cite{PR1091492}. This provides the clue to get insight
into the mechanism of the above localization in future works. However, if the
noise becomes too large, it is beyond the scope of this mean-field model in this
work. The above result also hints that the topological phase can be reflected
by dynamics.

\section{Conclusion}
\label{CoC}
The parameter diagram and the dynamical features in terms of the appearance of topological
gapless edge modes have been explored for a topological superconductor model with periodic
and open boundary conditions in two orthogonal directions. The appearance of the gapless
edge modes can be reflected by the expectations of Pauli matrices in two global manners.
The dynamical feature of excitations behave distinctively in terms of the appearance of
gapless edge modes and related mostly to the states with proximate eigenenergy between the
pre- and post-quench Hamiltonians. The cusps in the return rate occurs until the parameter
passes the parameter boundary very slowly. The dynamical pattern of the excitation in the
region with gapless edge modes is different to that in the region without gapless edge modes.
The cusps behave robustly until the noise of the lattice leads to topological phase transition.
This work indicates one can detect topological phase in dynamical manners.
\section{Acknowledgements}
X.L. Zhao thanks National Natural Science Foundation of China, No.12005110, 11975132,
Natural Science Foundation of Shandong Province, China, No.ZR2020QA078, ZR2021LLZ001,
and Project of Shandong Province Higher Educational Science and Technology Program,
No.J18KZ012.

\end{document}